\documentclass[%
 aip,
 amsmath,amssymb,
 reprint,%
]{revtex4-1}

\usepackage{graphicx}
\usepackage{dcolumn}
\usepackage{bm}

\usepackage[utf8]{inputenc}
\usepackage[T1]{fontenc}
\usepackage{mathptmx}
\usepackage{etoolbox}
\usepackage{bigints}
\usepackage[dvipsnames]{xcolor}
\usepackage{comment} 
\usepackage{multirow}

\makeatletter
\def\@email#1#2{%
 \endgroup
 \patchcmd{\titleblock@produce}
  {\frontmatter@RRAPformat}
  {\frontmatter@RRAPformat{\produce@RRAP{*#1\href{mailto:#2}{#2}}}\frontmatter@RRAPformat}
  {}{}
}%
\makeatother
\begin{document}

\preprint{AIP/123-QED}

\title[Two-phase coexistence line of the THF hydrate]{Prediction of the univariant two-phase coexistence line of the tetrahydrofuran hydrate from computer simulation}

\author{Jesús Algaba}
\affiliation{Laboratorio de Simulaci\'on Molecular y Qu\'imica Computacional, CIQSO-Centro de Investigaci\'on en Qu\'imica Sostenible and Departamento de Ciencias Integradas, Universidad de Huelva, 21006 Huelva Spain}

\author{Crist\'obal Romero-Guzm\'an}
\affiliation{Laboratorio de Simulaci\'on Molecular y Qu\'imica Computacional, CIQSO-Centro de Investigaci\'on en Qu\'imica Sostenible and Departamento de Ciencias Integradas, Universidad de Huelva, 21006 Huelva Spain}

\author{Miguel J. Torrej\'on}
\affiliation{Laboratorio de Simulaci\'on Molecular y Qu\'imica Computacional, CIQSO-Centro de Investigaci\'on en Qu\'imica Sostenible and Departamento de Ciencias Integradas, Universidad de Huelva, 21006 Huelva Spain}

\author{F. J. Blas$^*$}
\email{felipe@uhu.es}
\affiliation{Laboratorio de Simulaci\'on Molecular y Qu\'imica Computacional, CIQSO-Centro de Investigaci\'on en Qu\'imica Sostenible and Departamento de Ciencias Integradas, Universidad de Huelva, 21006 Huelva Spain}

\begin{abstract}
In this work, the univariant two-phase coexistence line of the hydrate of tetrahydrofuran (THF) is determined from 100 to 1000 bar by molecular dynamics simulations. The study is carried out by putting in contact a THF hydrate phase with a stoichiometric aqueous solution phase. Following the direct coexistence technique, the pressure has been fixed, and the coexistence line has been determined by analyzing if the hydrate phase grows or melts at different values of temperature. The model of water used is the well-known TIP4P/Ice model. We have used two different models of THF based on the transferable parameters for phase equilibria-united atom approach (TraPPE-UA), the original (flexible) TraPPe-UA model as well as a rigid and planar version of it. Overall, at high pressures, small differences have been observed in the results obtained by both models. Also, large differences have been observed in the computational efforts required by the simulations performed using both models, being the rigid and planar version much faster than the original one. The effect of the unlike dispersive interactions between the water and THF molecules has been also analyzed at 250 bar using the rigid and planar THF model. In particular, we have modified the Berthelot combining rule by adding a factor ($\xi_\text{{O-THF}}$) that modifies the unlike water-THF dispersive interactions and we have analyzed the effect on the dissociation temperature when $\xi_\text{{O-THF}}$ is modified from 1.0 (original Berthelot combining rule) to 1.4 (modified Berthelot combining rule). We have extended the study using $\xi_\text{{O-THF}}=1.4$ and the rigid THF model to the rest of the pressures considered in this work, finding an excellent agreement with the scarce experimental data taken from the literature.

\end{abstract}

\maketitle
$^*$Corresponding author: felipe@uhu.es

\section{Introduction}

Clathrate hydrates are crystalline inclusion compounds consisting of a network of hydrogen-bonded molecules (host) forming cages in which small and medium molecules (guests) are encapsulated under the appropriate thermodynamic conditions.~\cite{Sloan2008a,Ripmeester2022a} Clathrate hydrates are simply called hydrates when the host molecule is water (H$_2$O). Hydrates have been studied in the last few decades because of the capability of these compounds to encapsulate molecules of environmental, industrial, and energetic interest. Hydrates can be used as a source of energy since there are huge amounts of methane (CH$_4$) as hydrate reservoirs on the ocean floors as well as on the permafrost.~\cite{lee2001methane,ruppel2017interaction} They can be also used for capturing greenhouse gases, such as carbon dioxide (CO$_2$),~\cite{ma2016review,dashti2015recent,cannone2021review,duc2007co2,choi2022effective,lee2014quantitative} as a secure and clean medium for nitrogen (N$_2$) recovery from industrial emissions,~\cite{Yi2019,hassanpouryouzband2018co2} and as alternative and safe solid structures for hydrogen (H$_2$) storage.~\cite{veluswamy2014hydrogen,lee2005tuning}  

Nature and level of occupancy of guest molecules in a hydrate have a huge impact on the stability conditions of these compounds as well as on the crystalline structure adopted by the hydrate.~\cite{Sloan2008a} Hydrates of small molecules, such as CO$_2$ or CH$_4$, crystallize in the so-called sI structure. The unit cell of the sI structure is formed from $46$ water molecules distributed in $6$ T (tetrakaidocahedron or $5^{12}6^{2}$) cages and $2$ D (pentagonal dodecahedron or $5^{12}$) cages, usually denoted as ``small'' and ``large'' hydrate cages. Hydrates of medium molecules, such as iso-butane, propane, and cyclopentane, crystallize in sII structure. The sII unit cell is more complex than the sI structure. This unit cell is formed from $136$ water molecules distributed in $16$ D (pentagonal dodecahedron or $5^{12}$) cages and $8$ H (hexakaidecahedron or $5^{12}6^{4}$) cages. Notice that the D or ``small cages'' are the same in both structures, but the ``large cages'' (H) are larger in the sII structure, allowing them to accommodate inside larger molecules. The sII structure has the peculiarity that can be stabilized by medium or small molecules, such as H$_2$ or N$_2$ via multiple occupancy of the H cages.~\cite{Brumby2019a,Mao2002a,Mao2004a,Katsumasa2007a,Papadimitriou2016a,Liu2017a,Belosudov2016a}

It is possible to tune the stability and/or the speed-growth of the hydrates using additives.~\cite{Sloan2008a,Li2012a} Those additives that are able to increase the speed of formation of the hydrates are called kinetic hydrate promoters. On the other hand, when the additives increase the stability of the hydrates, they are called thermodynamic hydrate promoters. Among the thermodynamic hydrate promoters, tetrahydrofuran or THF,~\cite{Larsen1998a,Torre2015a,Makino2005a,Linga2008a,Daraboina2013a} a cyclic five-member ether,  has been widely used to increase the stability conditions of hydrates. When THF is used as an additive, it can increase drastically the stability of hydrates reducing the pressure at which hydrates are stable.~\cite{Florusse2004a,Lee2005a,Strobel2007a,veluswamy2014hydrogen} THF only occupies the H cages (5$^{12}$6$^4$) of the sII hydrate structure. The T cages (5$^{12}$) remain empty and can be occupied by other guest molecules of small size and molecular weight.

The phase diagram of the water + THF binary mixture is a fascinating example of beauty and complexity of phase behavior of a two-component system involving solid, liquid, and gas phases. At high temperatures, above $277\,\text{K}$, and low and moderate pressures, below $1000\,\text{bar}$, solid phases do not play any role. Nevertheless, the mixture shows a complex fluid phase behavior. This part of the phase diagram has been investigated by some of us in a previous work.~\cite{Miguez2015b} It exhibits type VI phase behavior according to the Scott and Konynenburg classification of fluid phase behavior.\cite{Scott1970a,Konynenburg1980a} Particularly, the mixture shows a continuous gas-liquid critical line running between the critical points of pure water and THF, and a liquid-liquid immiscibility region bounded below and above by a critical line which corresponds to the upper and lower critical solution temperatures. At low pressures, the liquid-liquid critical line ends at two critical end-points linked by a liquid-liquid-vapor line running from $314$ to $420\,\text{K}$, approximately (see Fig.~6 of the work of M\'{\i}guez \emph{et al.}~\cite{Miguez2015b} for further details). Undeniably, the most salient feature of the phase diagram of the mixture is the characteristic region of closed-loop liquid-liquid immiscibility exhibited by the system. The mixture is completely miscible at low temperatures, below the left-side liquid-liquid critical line running from the LCEP of the mixture up to high pressures, but also at high temperatures, above the right-side liquid-liquid critical line running from the UCEP of the mixture up to high pressures. At intermediate temperatures, inside the region of the phase diagram located between the liquid-liquid-vapor three-phase line and liquid-liquid critical line of the mixture, the system is immiscible. Note that the liquid-liquid immiscibility region disappears at a maximum in the pressure, usually called a hypercritical point.

At low temperatures, at which solid phases play a key role, THF is water-soluble. This makes THF an unusual and peculiar thermodynamic hydrate promoter: it is able to form a stable sII hydrate by itself. THF hydrate is stable, at temperatures below $277\,\text{K}$ and at atmospheric pressure conditions.~\cite{Makino2005a}  It is important to remark that these are very mild conditions since most hydrates are stable at pressures above several times the atmospheric pressure.~\cite{Sloan2008a} Particularly, the THF hydrate exhibits three characteristic phase-equilibrium curves.~\cite{Makino2005a} Two of them are the usual hydrate -- aqueous solution -- gas three-phase coexistence curves. In one of them, the aqueous solution in equilibrium with the hydrate and gas phases has a THF composition below the stoichiometric ratio of the sII THF hydrate, $17$ molecules of water per each THF molecule (1 THF : 17 H$_2$O or $x_{\text{THF}}=0.0556$). In the second one, the composition of THF in the aqueous phase is greater than the stoichiometric ratio. The third one is an univariant hydrate -- aqueous solution two-phase coexistence curve. The existence of this univariant two-phase coexistence curve is one of the most characteristic points in the THF hydrate system.~\cite{Makino2005a} In fact, the composition of the aqueous solution in equilibrium with the sII THF hydrate is equal to the stoichiometric ratio of the sII THF hydrate (1:17 or $x_{\text{THF}}=0.0556$). This univariant curve of two-phase equilibria, as well as the two three-phase coexistence curves, converge at an invariant point at $277.45\,\text{K}$ and $3.9\,\text{kPa}$ under the stoichiometric composition.~\cite{Makino2005a} In this work, we concentrate on the determination of the univariant two-phase coexistence curve of the THF hydrate from computer simulation.

According to the main peculiarity of the THF hydrate, it is formed when THF is mixed in a stoichiometric ratio with water.~\cite{Asadi2019a,Sun2017a,Suzuki2011a,Sabase2009a} In other words, the concentration of THF in the aqueous solution is the same as that in the hydrate. Although most experiments are performed at stoichiometric conditions, it has been reported in the literature that the THF hydrate can form at higher and lower THF concentrations in the aqueous solution phase.~\cite{Strauch2018a,Ganji2006a,Andersson1996a,Chong2016a,Kumar2010a} This is special interesting for several reasons: (1) Even if the hydrate can form from a non-stoichiometric aqueous solution, it has been reported in the literature\cite{Strauch2018a} upper (82.7 wt\%) and lower (5.0 wt\%) limits for the THF concentration at which the hydrate can be formed. (2) When the THF hydrate is formed from an aqueous solution with a THF concentration below the stoichiometric one, the THF concentration in the solution is stabilized at a THF concentration of 6.5 wt\% approximately.~\cite{Liu2022a} Contrary, when the THF hydrate is formed from an aqueous solution with a THF concentration above the stoichiometric one, the THF concentration in the solution is stabilized at a THF concentration of 44.0 wt\% approximately.~\cite{Liu2022a} (3) Even if the THF hydrate can be formed from a non-stoichiometric aqueous solution, the speed of formation is higher at the stoichiometric conditions.~\cite{Liu2022a} (4) The THF hydrate formed from a non-stoichiometric aqueous solution is always a stoichiometric one, it means that there is always a THF molecule inside each hexakaidecahedron or H cage.~\cite{Strauch2018a,Ganji2006a,Andersson1996a,Chong2016a,Kumar2010a}

Although THF has been widely used as a hydrate promoter,\cite{Larsen1998a,Torre2015a,Makino2005a,Linga2008a,Daraboina2013a,Florusse2004a,Lee2005a,Strobel2007a,veluswamy2014hydrogen} there is a lack of studies about this compound from a molecular perspective. In a series of works, some of us studied the phase equilibria and interfacial properties of THF\cite{Garrido2016a} and its mixtures with CO$_2$, ~\cite{Garrido2017a,Algaba2018a,Miguez2015b} CH$_4$,~\cite{Miguez2015b,Algaba2019a} and water\cite{Miguez2015b} from theory, experiments, and molecular dynamic simulation. In these works, we proposed a new model of THF based on the widely-known TraPPE-UA (Transferable Potentials for Phase Equilibria - United Atoms) parametrization force field.~\cite{Keasler2012a} This model is a rigid and planar version of the original TraPPE-UA THF model, where bending and torsional degrees of freedom are frozen. We demonstrated that the rigid and planar version of the THF model predicts the same coexistence phase diagram and interfacial properties as the original flexible version.~\cite{Garrido2016a,Algaba2018a,Algaba2019a} Besides, the rigid version is about ten times faster than the original flexible one. 
As there are no internal degrees of freedom to consider, larger simulation time steps can be employed, resulting in shorter required simulation times.~\cite{Algaba2018a} Since long simulation times are required to study hydrates, it is necessary to use optimized and fast-to-simulate molecular models. That makes the rigid TraPPE-UA THF model an ideal candidate for studying hydrates of this thermodynamic promoter. Nevertheless, flexibility can affect how the THF molecules behave when they are encapsulated inside the hydrate. According to this, it is also necessary to analyze the dissociation line of the THF hydrate using the original TraPPE-UA flexible model.

A molecular understanding of the THF hydrate is crucial to understand how this additive can be efficiently used as a hydrate promoter to improve their capabilities for capturing CO$_2$, storing H$_2$, recovering N$_2$ or as CH$_4$ reservoir, among others. There are only a few experimental studies devoted to the phase diagram determination of the THF hydrate,~\cite{Sloan2008a,Makino2005a,Dyadin1973a,Manakov2003a} and to the best of our knowledge, this is the first time the THF hydrate is studied from a molecular perspective. The main objective of this work is to determine the univariant two-phase dissociation temperature of the THF hydrate, at several pressures, combining accurate molecular models for water and tetrahydrofuran and the direct computer simulation technique. In addition to this, we also analyze the effect of the THF flexibility on the stability of the hydrate.

The organization of this paper is as follows: In Sec. II, we describe the molecular models and the simulation details used in this work. The results obtained, as well as their discussion, are described in Sec. III. Finally, conclusions are presented in Sec. IV

\section{Molecular Models and Simulation Details}
\subsection{Molecular Models}

In this work, THF has been modeled using the flexible TraPPE-UA THF model proposed by Keasler~\emph{et al.}~\cite{Keasler2012a} as well as a rigid and planar version of this model proposed by some of us in previous papers.~\cite{Garrido2016a,Algaba2018a,Algaba2019a} In both models, THF is described by three different types of united-atoms interaction centers: the oxygen of the ether group (O), two $\alpha$-CH$_2$ methyl groups bonded directly to the oxygen ether group, and, closing the ring, two $\beta$-CH$_2$ methyl groups bonded between them and to the $\alpha$-CH$_2$ methyl groups. The parameters that describe the non-bonded interactions, as well as the partial charges located at each interaction center, of the rigid THF model are identical to those used in the original TraPPE-UA model.~\cite{Keasler2012a} However, bending and torsional degrees of freedom are frozen and angles have been fixed to their equilibrium values. As a consequence, although the angles have been fixed to the equilibrium values of the original model, in order to keep planar the molecule, the bond lengths had to be slightly modified in order to accommodate all the interaction centers in the same plane while the fixed bonded angles remain constant (we refer the reader to our previous works\cite{Garrido2016a,Algaba2018a,Algaba2019a} for further details). Water molecule is described using the well-known TIP4P/Ice model.~\cite{Abascal2005b,Conde2017a} 

In this work, the classical Lorentz combining rule has been applied in order to calculate the parameters for the unlike-size interactions between different groups. Also, the Berthelot combining rule between THF and water groups has been modified in order to match the experimental dissociation line of the THF hydrate:~\cite{Makino2005a}

\begin{equation}
    \epsilon_{\text{O-THF}}=\xi_\text{{O-THF}}(\epsilon_{\text{OO}}\,\epsilon_\text{{THF-THF}})^{1/2}
\end{equation}

\noindent where $\epsilon_{\text{O-THF}}$ is the well depth associated with the LJ potential for the unlike interactions between the oxygen of water molecule, O, and THF-groups, $\epsilon_\text{{OO}}$ and $\epsilon_\text{{THF-THF}}$ are the well depth for the like interactions between water-O and THF-groups respectively and $\xi_\text{{O-THF}}$ is the factor that modifies the Berthelot combining rule. In this work, different values of $\xi_\text{{O-THF}}$ have been used from 1.0 to 1.4 (see Section III for more details).

\subsection{Simulation Details}

All the results presented in this work have been obtained from molecular dynamics simulation using the GROMACS package (version 4.6 double-precision). Simulations have been carried out using the Verlet-leapfrog algorithm with a time step of $2\,\text{fs}$ for solving Newton's equations of movement when the rigid THF model is used. Notice that for the case of the original flexible THF model, a time step of $1\,\text{fs}$ was used in order to take into account correctly the bending and torsional degrees of freedom.

The dissociation line of the THF hydrate has been determined using the direct coexistence technique. Note that in the case of the THF hydrate, the dissociation line is a locus of two-coexisting phases, the hydrate phase and the aqueous phase. We denote the corresponding dissociation temperature, at a given pressure, as $T_{2}$. Following the direct coexistence methodology, a stochiometric aqueous solution phase of THF is put in contact via a planar interface with a THF hydrate phase in the same simulation box. By varying the temperature, it is possible to calculate the dissociation temperature $T_{2}$ of the hydrate, at a given pressure. If the temperature is above the equilibrium temperature, $T>T_{2}$, the hydrate melts. If the temperature is below the equilibrium temperature, $T<T_{2}$, the hydrate grows. According to this, the dissociating temperature is between the highest value of temperature at which the hydrate grows and the lowest at which the hydrate melts.

\begin{figure}
\includegraphics[width=\columnwidth]{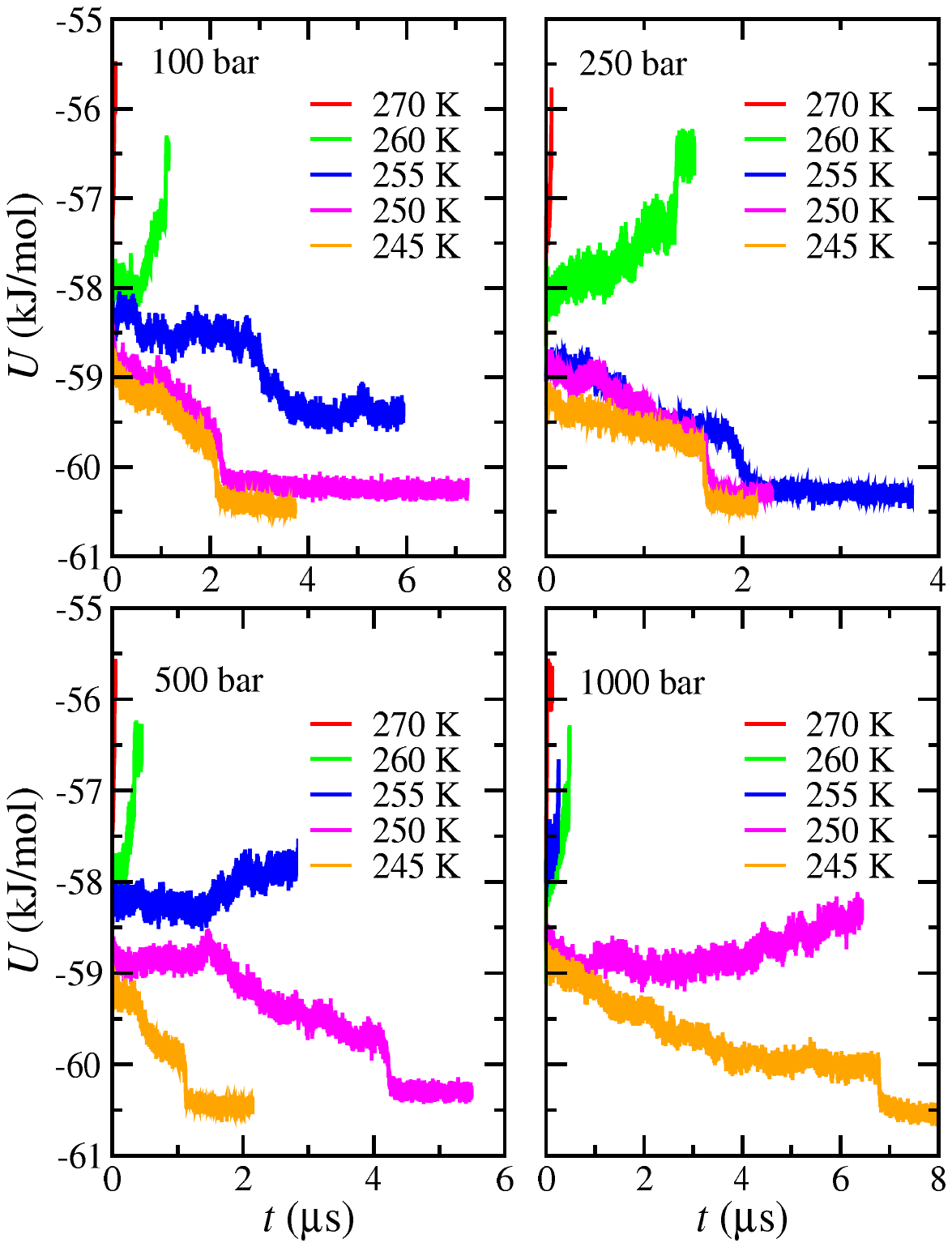}\\
\caption{Evolution of the potential energy of the THF hydrate -- aqueous solution configuration, as a function of time, as obtained from MD $NPT$ simulations at $100$, $250$, $500$, and $1000\,\text{bar}$ and several temperatures. Water and THF molecules are modeled using the TIP4P/Ice force field and the rigid version of the TraPPE model, respectively. In all cases, we use the Berthelot combining rule for dispersive interactions ($\xi_\text{{O-THF}}=1.0$).}
\label{figure1}
\end{figure}

\begin{figure}
\includegraphics[width=\columnwidth]{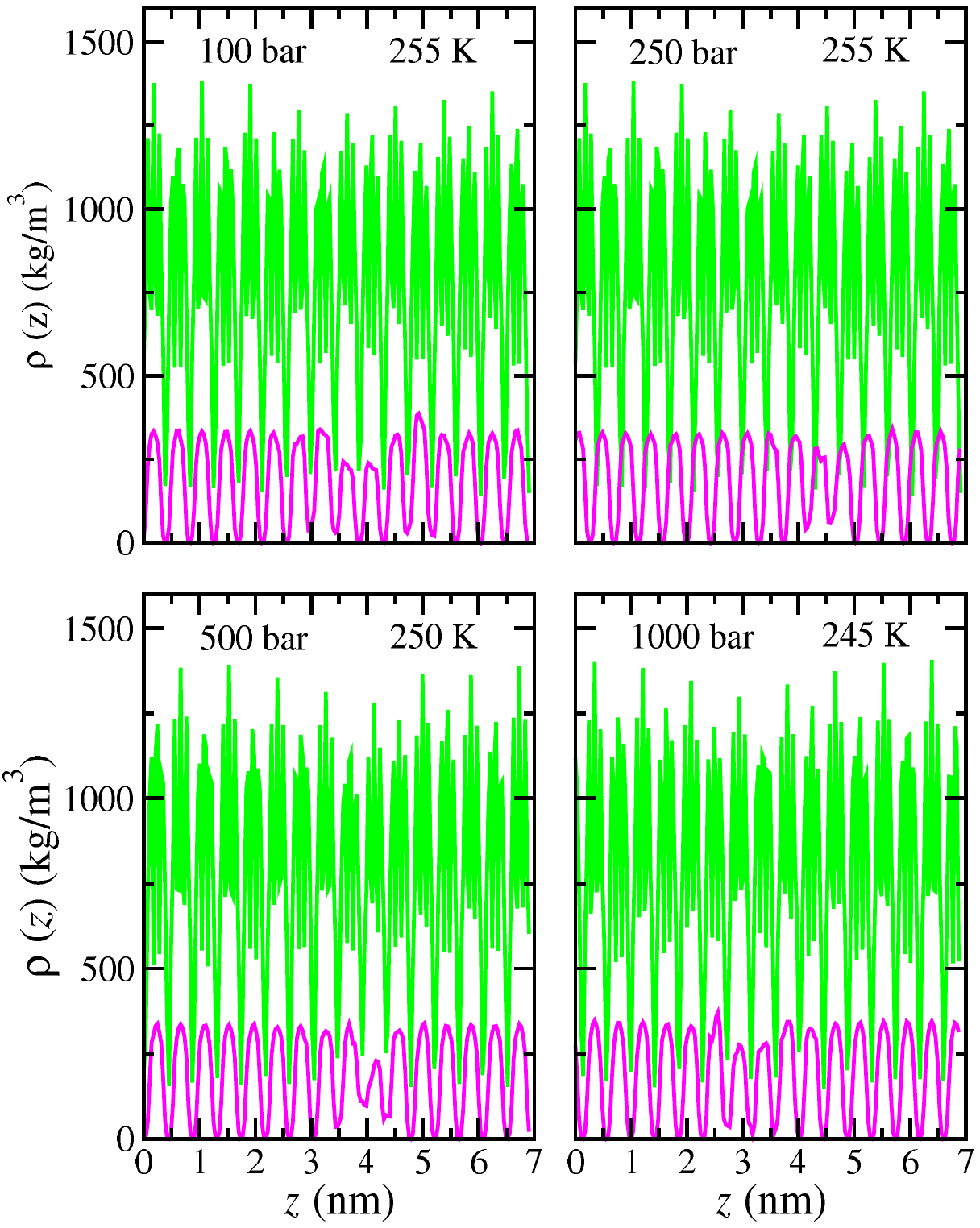}\\
\caption{Simulated equilibrium density profiles, $\rho(z)$, across the hydrate~--~aqueous solution interface of THF (magenta) and water (green) as obtained from MD $NPT$ simulations at several temperatures and pressures. Water and THF molecules are modeled using the TIP4P/Ice force field and the rigid version of the TraPPE model, respectively. In all cases, we use the Berthelot combining rule for dispersive interactions ($\xi_\text{{O-THF}}=1.0$).}
\label{figure2}
\end{figure}

In all cases, the hydrate phase is built by replicating twice the THF hydrate unit cell in the three space directions ($2\times2\times2$). The initial  THF hydrate phase is formed from $1088$ molecules of water and $64$ molecules of THF. The same number of molecules is used to build the aqueous solution phase in contact with the hydrate phase. The interface between both phases is arbitrarily placed perpendicular to the $z$ direction of the simulation box. The initial size of the simulation box is the same in all cases: $L_x=L_y=3.47$ nm and $L_z=6.80$ nm.

In order to keep constant the pressure and temperature, simulations are carried out in the isobaric-isothermal $NPT$ ensemble. The Parrinello-Rahman barostat\cite{Parrinello1981a} has been used with a time constant of 1 ps and a compressibility value of 4.5e$^{-5}$. In order to avoid stress from the solid structure, the Parrinello-Rahman barostat is applied independently in the three directions of the simulation box instead of only in the direction perpendicular to the interface. The V-rescale thermostat algorithm\cite{Bussi2007a} with a time constant of 0.05 ps is chosen to fix the temperature value along the simulation. Non-bonded Lennard-Jones and coulombic interactions are truncated by a 1.55 nm cut-off. No long-range corrections are used for the Lennard-Jones dispersive interactions and particle-mesh Ewald (PME)\cite{Essmann1995a} corrections are applied for the coulombic potential.

\begin{table}
\centering
\begin{tabular}{ccccccc}
\hline\hline
\multirow{2}{*}{$P$ (bar)} & &$T_2^{flex}$ (K) & &$T_2^{rigid}$ (K) & & $T_2^{rigid}$ (K)\\
& & $\xi_\text{{O-THF}}=1.0$ & & $\xi_\text{{O-THF}}=1.0$ & & $\xi_\text{{O-THF}}=1.4$ \\
\hline
100 & &258(3) & &258(3)& & 278(3)\\
250 & & 258(3) & &258(3)& & 278(3)\\
500 & &258(3) & &253(3)& & 273(3)\\
1000 & &253(3) & &248(3)& & 268(3)\\
\hline\hline
\end{tabular}
\caption{Dissociation temperature, $T_{2}$, of the THF hydrate, as a function of pressure $P$, along the univariant THF hydrate -- aqueous solution two-phase coexistence curve. Water molecules are modeled using the TIP4P/Ice force field. The different dissociation temperatures correspond to the predictions using the flexible and the rigid version of the TraPPE model and different values of $\xi_\text{{O-THF}}$.}
\label{table-T2}
\end{table}

 \section{Results}
In this section, we show the results corresponding to the dissociation line of the THF hydrate obtained from molecular dynamics simulations. First, we consider the effect of pressure on the dissociation temperature $T_{2}$. Using the rigid THF model and $\xi_\text{{O-THF}}=1.0$, the $T_2$ is determined at four pressures: $100$, $250$, $500$, and $1000\,\text{bar}$. Secondly, the effect of the flexibility of the THF model on the $T_{2}$, at the same pressures, is analyzed using the original and flexible TraPPE-UA THF model. Thirdly, the effect of different values of $\xi_\text{{O-THF}}$ (from 1.0 to 1.4) on the $T_{2}$ is analyzed at $250\,\text{bar}$ using the rigid and planar THF models. Finally, the two-phase dissociation line is studied using the rigid and planar THF model with the optimized value of $\xi_\text{{O-THF}}=1.4$.

\subsection{Two-phase line of the hydrate. Rigid model for THF}

We first focus on the results obtained using $\xi_\text{{O-THF}}=1.0$ and the rigid and planar THF model proposed by some of us in a previous series of papers.~\cite{Garrido2016a,Algaba2018a,Algaba2019a}As we have mentioned previously, this model is able to predict accurately the phase equilibria and interfacial properties of pure THF\cite{Garrido2016a} and its binary mixtures with CO$_{2}$\cite{Algaba2018a} and CH$_{4}$,~\cite{Algaba2019a} which exhibit vapor-liquid and liquid-liquid phase behavior, respectively.

Fig.~\ref{figure1} shows the evolution of the potential energy, as a function of simulation time, of the THF hydrate--aqueous solution system at four different pressures. As it has been explained previously and according to the direct-coexistence methodology, the dissociation temperature $T_{2}$ can be obtained by performing simulations at different temperatures and analyzing if the solid phase melts or grows. If the hydrate melts/grows, the simulated temperature is above/below the dissociation temperature $T_{2}$. When the hydrate grows, the potential energy of the system increases (in absolute value). This increase of potential energy, in absolute value, is related to the increase in the formation of new hydrogen bonds due to the crystallization of the aqueous solution phase. Contrary, when the hydrate phase melts, the absolute value of the potential energy decreases. The $T_{2}$ is in the middle between the highest temperature at which the hydrate grows and the lowest temperature at which the hydrate melts. According to this, it is possible to determine the dissociation temperature. From this analysis, we conclude that the dissociation temperatures at $100$, $250$, $500$, and $1000\,\text{bar}$ using $\xi_{\text{O-THF}}=1.0$ and the rigid and planar THF model are $258(3)$, $258(3)$, $253(3)$, and $248(3)\,\text{K}$ respectively. All the results are summarized in Table \ref{table-T2}.

We have also determined the equilibrium density profiles of the configurations that crystallize at the highest temperature but below the estimated $T_{2}$ and results are presented in Fig.~\ref{figure2}. The inspection of the density profiles provides a complementary method to check if the hydrate crystallizes, and hence, to accurately locate the dissociation temperature of the hydrate at a given pressure. As can be seen in Fig.~\ref{figure2}, in all cases the density profiles of the THF and H$_{2}$O molecules show the characteristic peaks of the hydrate crystal structure. Contrary, if the system melts, density profiles exhibit the classical flat behavior of a homogeneous aqueous solution (not shown here).

\begin{figure}
\includegraphics[width=\columnwidth]{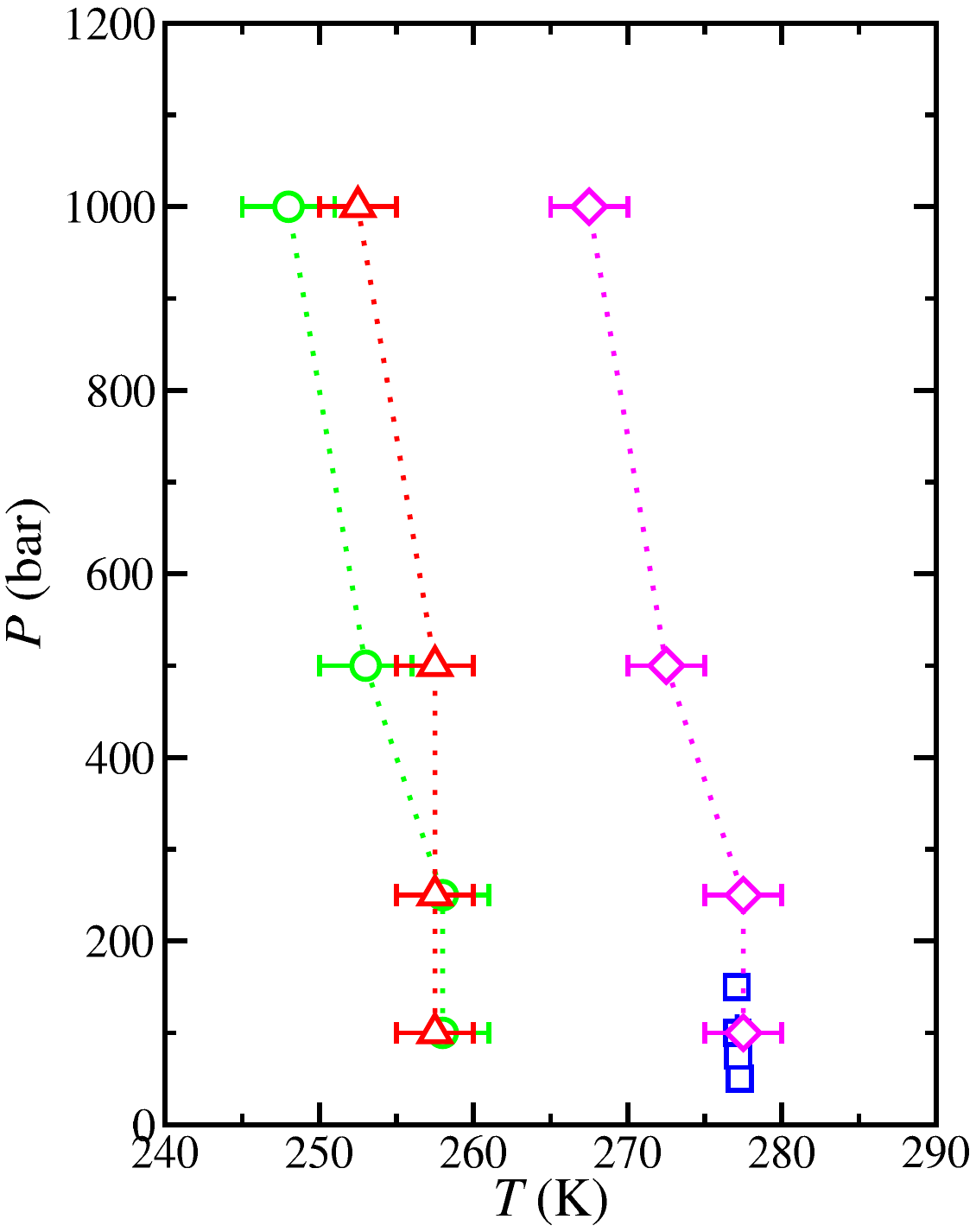}\\
\caption{Pressure-temperature projection of the dissociation line of the THF hydrate as obtained from MD $NPT$ simulations. Water and THF molecules are modeled using the TIP4P/Ice force field and the TraPPE models, respectively. Green circles and red triangles represent the results obtained using the Berthelot combining rule for dispersive interactions ($\xi_\text{{O-THF}}=1$) using the rigid and flexible THF models, respectively. Magenta diamonds represent the results obtained using the optimized unlike dispersive interactions ($\xi_\text{{O-THF}}=1.4$) and the rigid THF TraPPE model. Blue squares represent the experimental data taken from the literature.~\cite{Sloan2008a}}
\label{figure3}
\end{figure}

We also present the pressure-temperature projection of the univariant two-phase line of the THF hydrate. Fig.~\ref{figure3} shows the $T_{2}$ values, at the corresponding pressures, of the THF hydrate using the rigid and planar model of THF (green circles) and the original Berthelot rule for the unlike dispersive interactions between water and THF ($\xi_{\text{O-THF}}=1.0$). As we have already mentioned, the dissociation temperature of the system is unaffected at low pressures, and only for pressures above $500\,\text{bar}$ the $T_{2}$ slightly decreases when the pressure is increased. In other words, the pressure only affects the hydrate stability at very high pressures. We have also represented in the same figure the experimental $T_{2}$ values taken from the literature (blue squares).~\cite{Sloan2008a,Makino2005a} As can be seen, computer simulation predictions obtained using the rigid model of THF and $\xi_{\text{O-THF}}=1.0$ underestimate around $20\,\text{K}$ the experimental $T_{2}$ values in the whole range of pressures.

As we have already mentioned, the dissociation temperature at $100$ and $250\,\text{bar}$ is the same within the error bars, indicating that pressure has little or no effect on $T_{2}$. As we can be see in Figs.~\ref{figure1} and \ref{figure3} and in Table~\ref{table-T2}, $T_{2}=258(3)\,\text{K}$ at both pressures. At higher pressures, from $500$ to $1000\,\text{bar}$, the effect of pressure on the $T_{2}$ is increased, displacing the dissociation temperature from $253(3)$, at $500\,\text{bar}$, to $248(3)\,\text{K}$ at $1000\,\text{bar}$. When the pressure is increased, the dissociation temperature decreases. According to this, the THF hydrate loses stability as the pressure increases. Probably, this effect is related to the hydrate cages' size. Each THF molecule occupies one hexakaidecahedron or H (large) cage of the hydrate structure because of its molecular size. When the pressure is increased, this has a little but non-negligible effect on the hydrate structure, reducing the size of the cage voids and making the occupancy of large molecules less stable.

\subsection{Effect of flexibility on $T_{2}$}

The effect of the flexibility of the THF model on the $T_{2}$ is also analyzed in this work. Once the dissociation line is determined using $\xi_\text{{O-THF}}=1.0$ and the rigid version of the THF model, extra simulations are carried out using the original and flexible TraPPE-UA THF model.~\cite{Keasler2012a} The flexible THF model requires larger computational efforts than the planar and rigid version.~\cite{Algaba2018a} Due to this, we only study two initial temperatures at each pressure: the highest temperature at which the hydrate grows and the lowest temperature at which the hydrate melts using the rigid and planar THF model (see Fig.~\ref{figure1} for further details). Note that we use the same $\xi_\text{{O-THF}}$ value.

\begin{figure}
\includegraphics[width=\columnwidth]{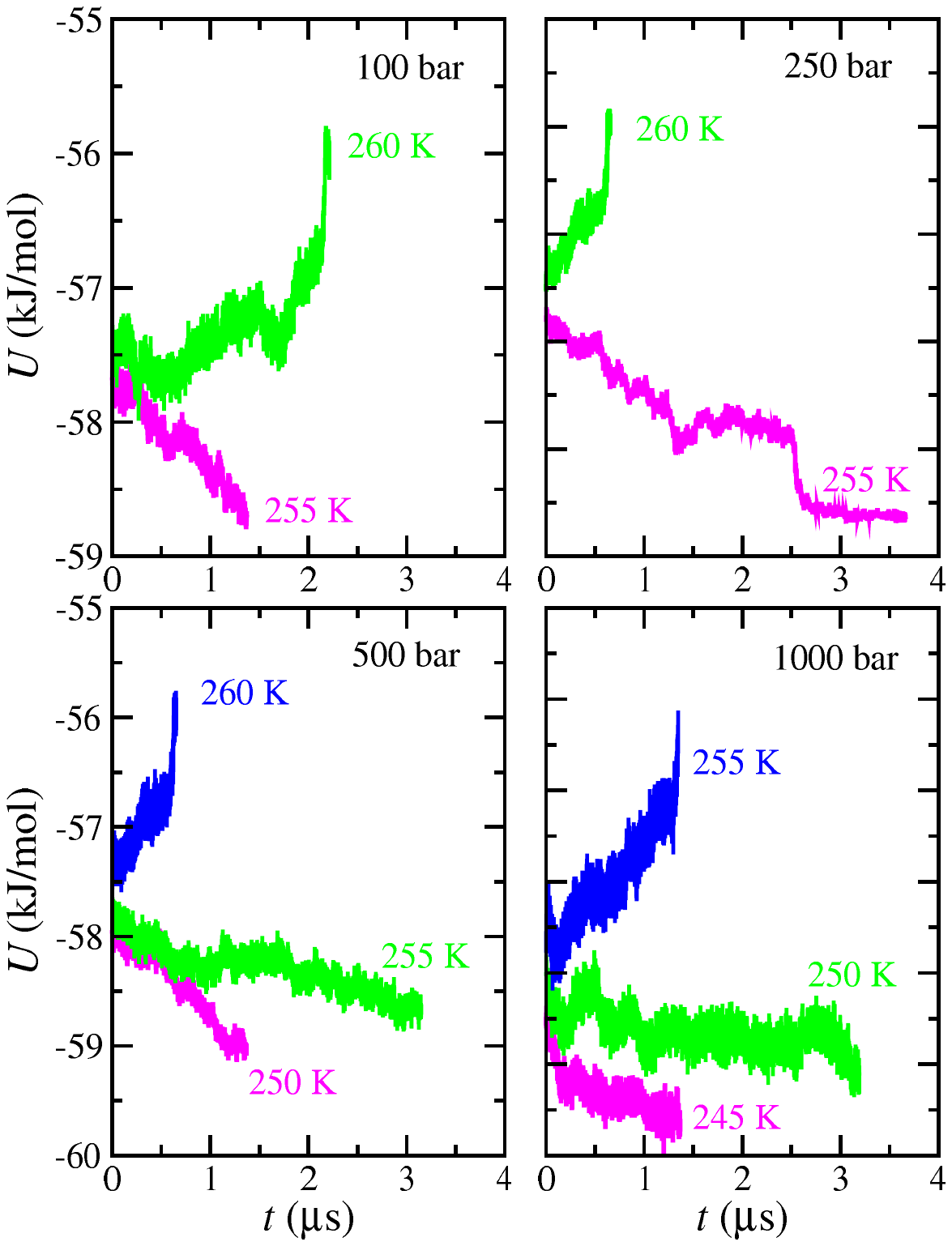}\\
\caption{Evolution of the potential energy of the THF hydrate -- aqueous solution configuration, as a function of time, as obtained from MD $NPT$ simulations at $100$, $250$, $500$, and $1000\,\text{bar}$ and several temperatures. Water and THF molecules are modeled using the TIP4P/Ice force field and the flexible version of the TraPPE model, respectively. In all cases, we use the Berthelot combining rule for dispersive interactions ($\xi_\text{{O-THF}}=1$).}
\label{figure4}
\end{figure}

Results for the evolution of the potential energy of the system using the TIP4P/Ice force field for water and the flexible version of the TraPPE model for THF are presented in Fig.~\ref{figure4}. As can be seen, the $T_{2}$ values at $100$ and $250\,\text{bar}$ obtained using the flexible model of THF are the same as those determined using the rigid model. However, at $500$ and $1000\,\text{bar}$, the hydrate phase crystallizes at a higher temperatures than those corresponding to the rigid and planar THF TraPPE version. In order to accurately determine the dissociation temperatures at both pressures, we simulate the system at other temperatures, as it is shown in Fig.~\ref{figure4}. Following a similar analysis as in Section III.A., the dissociation temperatures at $500$ and $1000\,\text{bar}$ are $258(3)$ and $253(5)\,\text{K}$, respectively. All the results obtained in this work are summarized in Table \ref{table-T2}.

We also show the predictions obtained for the flexible model of THF in the pressure-temperature projection in Fig.~\ref{figure3}. Results obtained using the flexible model are represented now using red-up triangles. As can be seen, comparing the dissociation lines obtained using the rigid and flexible models of THF, flexibility has little or negligible effects on $T_{2}$ at low pressures, as we have previously discussed. However, as the pressure increases, the dissociation line of the flexible model is displaced towards higher temperatures with respect to that of the rigid model. In other words, at high pressures, the THF hydrate modeled using a flexible force field is more stable than when a rigid one is used. Why? The original TraPPE-UA model of THF is flexible. Due to this, the flexible THF molecules can be accommodated inside the hexakaidecahedron or H (large) cages more efficiently than the rigid and planar molecules. In terms of relative stability, the crystalline sII solid phase, when the flexible THF molecules occupy the large cages of the structure, is more stable than when the rigid THF molecules are enclathrated. Although simple, we believe this argument could elucidate why the stability of the THF hydrate modeled with the original flexible TraPPE-UA model slightly surpasses that with the rigid molecules at elevated pressures.

The recap, there exist small differences in the dissociation line of the THF hydrate when using the rigid and flexible model. At low pressures, the dissociation temperature of both models is the same. At higher pressures, however, we observe small differences, although the $T_{2}$ values are equal within the error bars. This is in good agreement with the results obtained by some of us in previous papers, where the interfacial and phase equilibria of pure THF\cite{Garrido2016a} and THF+CO$_2$ binary mixture\cite{Algaba2018a} were studied using both models without finding significant differences in the results. However, it is necessary to take into account that we are using the direct coexistence simulation technique. According to this, when the temperature of the system is close to the $T_{2}$, the system can grow even when the temperature is above the $T_2$ and can melt even when the temperature is above the $T_2$. This is due to the inherent thermal fluctuations and the stochasticity of the methodology. In order to increase the accuracy of the results, it would be desirable to run more than one seed at each simulated temperature close to the $T_{2}$.~\cite{Conde2017a} However, due to the long simulation times required in this work, running more than one seed was not feasible. So it is not possible to ensure that the small differences observed in the $T_{2}$ of both models come from the effect of the flexibility at high pressures, from the limitations of the direct coexistence technique, or from both.


\subsection{Effect of dispersive interactions on $T_{2}$}

In Section III.B., we have demonstrated that the rigid model presented by us several years ago~\cite{Garrido2016a,Algaba2018a,Algaba2019a} is able to provide similar results than the more realistic original TraPPE model.~\cite{Keasler2012a} However, the rigid model, which does not account for the bending and torsional internal degrees of freedom of the molecule, is a better option than the flexible one in terms of CPU time. Due to these reasons, we only concentrate from this point on the rigid model of THF. Unfortunately, none of the models are able to provide a quantitative description of the experimental two-phase coexistence line of the THF hydrate (see Fig.~\ref{figure3}). The agreement between simulation predictions and experimental data taken from the literature can be improved by modifying the Berthelot combining rule. This allows to tune the unlike dispersive interactions between the THF chemical groups and the oxygen atom of the water molecule in order to get the best description of the experimental data. This has been previously done with success by several authors to quantitatively describe the dissociation lines of the CO$_{2}$~\cite{Miguez2015a,Constandy2015a,Waage2017a,Algaba2023a,Algaba2024a,Algaba2024b},  N$_{2}$,~\cite{Algaba2023b} and H$_2$.~\cite{Michalis2022a}

As far as the authors know, there is a lack of information in the literature about the phase diagram of the THF hydrate.~\cite{Sloan2008a,Makino2005a,Dyadin1973a,Manakov2003a} Most experimental data available in the literature corresponds to measures at low pressures, below $200\,\text{bar}$. There is another experimental point, at high pressures, that corresponds to a hydrate structural transition point from the sII to sI structure of the solid phase ($2000\,\text{bar}$ and $268\,\text{K}$).~\cite{Dyadin1973a,Manakov2003a} The experimental data taken from the literature has been included in the pressure-temperature projection of the dissociation line of the TFH hydrate shown in Fig.~\ref{figure3}. 

In this section, we analyze the effect of $\xi_\text{{O-THF}}$ on the $T_{2}$ value at $250\,\text{bar}$. We perform the study at this pressure mainly for two reasons. Firstly, although there is no experimental data at $250\,\text{bar}$, the dissociation temperature obtained from simulations at $100$ and $250\,\text{bar}$ is the same. This is in good agreement with the work carried out by Makino~\emph{et al.}~\cite{Makino2005a} In fact, these authors have demonstrated that the univariant THF hydrate -- aqueous solution two-phase line shows a slope very steep in the pressure-temperature diagram at low pressures, i.e., $T_2$ is independent of the pressure. This is clearly shown in Fig.~\ref{figure3}, where the green circles represent the predictions using $\xi_\text{{O-THF}}=1.0$ and the blue squares represent the experimental data taken from the literature. Secondly, we expect the same behavior of the predicted dissociation line at $100$ and $250\,\text{bar}$ using higher values of $\xi_\text{{O-THF}}$. Since $250\,\text{bar}$ is closer to the high-pressure region of the dissociation line, the optimized value for $\xi_\text{{O-THF}}$ would provide a representative prediction of the system behavior at higher pressures.

We have determined the hydrate dissociation temperature, at $250\,\text{bar}$, for different $\xi_\text{{O-THF}}$ values (1.1, 1.3, and 1.4). The results are summarized in Table \ref{table-T2-chi}. As can be observed, the $T_{2}$ values increase as $\xi_{\text{O-THF}}$ values are larger, from 1.0 to 1.4. This is an expected behavior also observed in other hydrates in the literature.\cite{Miguez2015a,Michalis2022a,Algaba2023b} Note that the increment is not linear. In fact, the variation of $T_{2}$ is larger when $\xi_\text{{O-THF}}$ varies from $1.0$ to $1.1$ than when is increased from $1.3$ to $1.4$. Actually, the $T_{2}$ values obtained using $\xi_{\text{O-THF}}=1.3$ and 1.4 are almost the same since the error bars of both temperatures overlap (see Table~\ref{table-T2} and Fig.~\ref{figure3}). Taking into account that $T_2$ is the same at $100$ and $250\,\text{bar}$ independently of the $\xi_{\text{O-THF}}$ value (see the next section), the result obtained when $\xi_{\text{O-THF}}=1.4$ is in excellent agreement with the experimental data taken from the literature~\cite{Sloan2008a} at $100\,\text{bar}$ (see Fig.~\ref{figure3}). In the next section, we use this optimized $\xi_{\text{O-THF}}$ value in a transferable way and obtain the dissociation temperature of the hydrate at other pressures.

\begin{table}
\centering
\begin{tabular}{ccccccccc}
\hline\hline

$\xi_\text{{O-THF}}$ & & $1.0$ & & $1.1$ & & $1.3$ & & $1.4$ \\
\\
$T_{2}\,\text{(K)}$ & & 258(3) & & 265(5)& &273(3) & & 278(3) \\
\hline\hline
\end{tabular}
\caption{Dissociation temperature, $T_2$, of the THF hydrate at $250\,\text{bar}$. Water molecules are modeled using the TIP4P/Ice force field. In all cases cases THF was modelled using the rigid version of the TraPPe mode. The different dissociation temperatures correspond to the different values of $\xi_\text{{O-THF}}$ employed.}
\label{table-T2-chi}
\end{table}

\subsection{Two-phase line of the hydrate. Optimized model for THF}

We now consider the univariant two-phase coexistence line of the THF hydrate using in a transferable way the $\xi_\text{{O-THF}}$ value obtained in Section III.C,  at $250\,\text{bar}$. To this end, we follow the same approach used in Section III. A and select two temperatures at each of the pressures considered, $100$, $500$, and $1000\,\text{bar}$, in addition to the $250\,\text{bar}$ pressure previously considered. 

\begin{figure}
\includegraphics[width=\columnwidth]{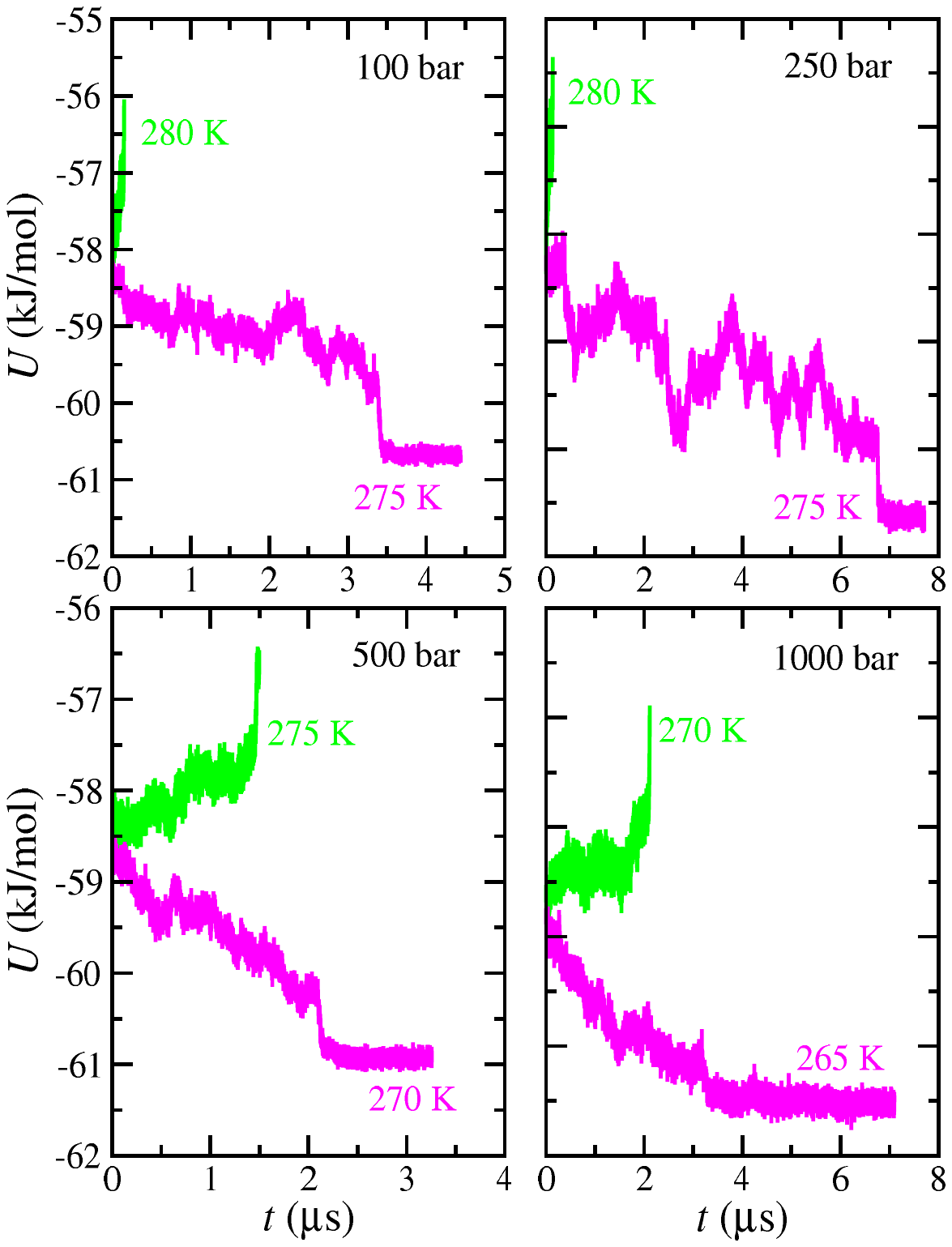}\\
\caption{Evolution of the potential energy of the THF hydrate -- aqueous solution configuration, as a function of time, as obtained from MD $NPT$ simulations at $100$, $250$, $500$, and $1000\,\text{bar}$ and several temperatures. Water and THF molecules are modeled using the TIP4P/Ice force field and the rigid version of the TraPPE model, respectively. In all cases, we use the optimized factor $\xi_\text{{O-THF}}=1.4$ for unlike dispersive interactions.}
\label{figure5}
\end{figure}

Fig.~\ref{figure5} shows the evolution of the potential energy, as a function of the simulation time, of the THF hydrate–aqueous solution system at four different pressures. Results are obtained using the TIP4P/Ice and rigid TraPPE models for water and THF, respectively. In all cases, $\xi_\text{{O-THF}}=1.4$. For each pressure, we simulate the highest temperature at which the hydrate grows (curves in magenta) and the lowest temperature at which the hydrate melts (curves in green). As can be seen, the behavior at $100\,\text{bar}$ is the expected one: the increase of $\xi_\text{{O-THF}}$ from $1.0$ to $1.4$ provokes an increment in the $T_{2}$ of $20\,\text{K}$, from $258(3)$ to $278(3)\,\text{K}$. This is in agreement with the discussion previously mentioned in Section III.C. The same effect can be observed at higher pressures: at $500$ and $1000\,\text{bar}$, the variation of $\xi_{\text{O-THF}}$ also provokes a displacement of $20\,\text{K}$ in the corresponding $T_{2}$ values. Particularly, from $253(3)$ to $273(3)\,\text{K}$ and from $248(3)$ to $268(3)\,\text{K}$ at $500$ and $1000\,\text{bar}$, respectively. A general picture of the effect of varying $\xi_{\text{O-THF}}$ can be observed in Fig.~\ref{figure3}. As can be seen, increasing $\xi_{\text{O-THF}}$ does not provoke a change in the qualitative behavior of the pressure-temperature projection of the univariant two-phase line since the whole $T_{2}$ line is shifted $20\,\text{K}$ in the diagram (magenta curve). 

To assess if predictions obtained from computer simulations of the proposed model are able to provide a quantitative description of the univariant two-phase line of the THF hydrate, we have also included experimental data taken from the literature\cite{Sloan2008a} in Fig.~\ref{figure3}. Unfortunately, as we have already mentioned, there is a lack of experimental data at moderate and high pressures. As a consequence of this, only simulation results obtained at $100\,\text{bar}$ can be directly compared with the experimental data taken from the literature. As can be observed in Fig.~\ref{figure3}, agreement between both results is excellent within the error bars of simulation data. It is interesting to remark that simulation results agree with the experimental findings at low pressures,~\cite{Sloan2008a,Makino2005a} i.e., both simulation and experimental $T_{2}$ values are independent of pressure. In addition to this, the dissociation temperature decreases as pressure is increased. This is also in agreement with experiments. According to them, there exists a structural transition point from sII to sI structure at $268\,\text{K}$ and $2000\,\text{bar}$.~\cite{Makino2005a,Dyadin1973a,Manakov2003a} Since there is a poor dependency of the dissociation temperature with pressure and simulation predictions indicate that $T_{2}=268(3)\,\text{K}$ at $1000\,\text{bar}$, we think the combination of the TIP4P/Ice and rigid TraPPE models for water and THF, with the unlike dispersive interactions proposed in this work, also provide confident predictions of the univariant two-phase line of the THF hydrate at high pressures.

\section{Conclusions}

We have determined the univariant two-phase coexistence line of the THF hydrate combining molecular dynamics and the direct coexistence simulation technique in a wide range of pressures, from $100$ to $1000\,\text{bar}$. The study is carried out using the TIP4P/Ice model for water and two different models for THF. The first model for THF is the original and flexible TraPPE-UA model proposed by Keasler \emph{et al.}~\cite{Keasler2012a} and the second one is a rigid and planar version of it proposed by some of us in previous works.~\cite{Garrido2016a,Algaba2018a,Algaba2019a}

We first predict the dissociation line of the THF hydrate using the rigid and planar model of THF and the Lorentz-Berthelot combining rules for the water-THF unlike dispersive interactions. According to the direct coexistence simulation technique, the dissociation temperatures or $T_{2}$ are obtained by inspecting the evolution of the potential energy of the THF hydrate-aqueous solution configurations as functions of time at different temperatures and pressures. The results are corroborated by calculating the density profiles of water and THF along the direction perpendicular to the planar interface separating both phases. Computer simulations predict that the univariant two-phase line of the THF hydrate shows a slope very steep in the pressure-temperature projection of the phase diagram in a wide range of pressures, in qualitative agreement with experimental evidences.~\cite{Makino2005a,Dyadin1973a,Manakov2003a} Predictions from molecular dynamics simulations are directly compared with experimental data taken from the literature. Unfortunately, only data below $200\,\text{bar}$ is available in the literature. At these conditions, only results obtained at $100\,\text{bar}$ can be compared with experimental data. In this case, simulation results predict a $T_{2}$ value $20\,\text{K}$ below the experimental data.

We also consider the more realistic and flexible TraPPE-UA model of Keasler \emph{et al.}~\cite{Keasler2012a} We follow the same methodology used to determine the dissociation line of the rigid THF model, we also calculate the location of the $T_{2}$ in the whole range of pressures. According to the simulations results, the effect of flexibility of the THF model on the univariant two-phase line of the THF hydrate is negligible at pressures below $250\,\text{bar}$, i.e., the rigid and flexible models provide the same $T_{2}$ values at low pressures. At higher pressures, the $T_{2}$ values predicted using the flexible THF model are slightly displaced towards higher temperatures, $\sim5\,\text{K}$ with respect to those obtained using the rigid version. Since the simulations performed using the flexible model are expensive and both models provide similar results, within the simulation error bars, we decided to concentrate on the rest of the study only using the rigid model for THF.

With the aim of improving the agreement between simulation results and experimental data, we modify the deviation from the Berthelot combining rule associated with the water-THF dispersive interaction, $\xi_\text{{O-THF}}$, to find the best possible description of the experimental $T_{2}$ values from computer simulations. To this end, we analyze the effect of dispersive interactions on the dissociation temperature of the hydrate at $250\,\text{bar}$. We consider four different values of $\xi_\text{{O-THF}}$, $1.0$, $1.1$, $1.3$, and $1.4$. According to the results, the main effect of increasing $\xi_\text{{O-THF}}$ is to shift the $T_{2}$ towards higher temperatures. We find that $\xi_\text{{O-THF}}=1.4$ value provides the best description of the hydrate experimental dissociation temperature.

Finally, we use the $\xi_\text{{O-THF}}=1.4$ value determined at $250\,\text{bar}$ in a transferable way and predict the whole dissociation line at lower and higher pressures. We find an excellent agreement between simulation and experiments at $100\,\text{bar}$. It is important to remark that the use of $\xi_\text{{O-THF}}=1.4$ displaces the whole THF hydrate dissociating line by $20\,\text{K}$. To the best of our knowledge, this is the first time the univariant two-phase coexistence line of the THF hydrate is predicted from computer simulation using simple but accurate models for water and THF. We expect that the results obtained in this work provide insightful information about the THF hydrate and help in the future to improve studies about phase equilibria of hydrates containing  THF as thermodynamic hydrate promoters.

\section*{Acknowledgements}
This work was funded by Ministerio de Ciencia e Innovaci\'on (Grant No.~PID2021-125081NB-I00), Junta de Andalucía (P20-00363), and Universidad de Huelva (P.O. FEDER UHU-1255522 and FEDER-UHU-202034), all four co-financed by EU FEDER funds. CR-G acknowledges the FPI Grant (Ref.~PRE2022-104950) from Ministerio de Ciencia e Innovaci\'on and Fondo Social Europeo Plus. MJT also acknowledges the research contract (Ref.~01/2022/38143) of Programa Investigo (Plan de Recuperaci\'on, Transformaci\'on y Resiliencia, Fondos NextGeneration EU) from Junta de Andaluc\'{\i}a (HU/INV/0004/2022). We greatly acknowledge RES resources at Picasso provided by The Supercomputing and Bioinnovation Center of the University of Malaga to FI-2024-1-0017.

\section*{AUTHORS DECLARATIONS}

\section*{Conflicts of interest}

The authors have no conflicts to disclose.

\section*{Data availability}

The data that support the findings of this study are available within the article.

\section*{REFERENCES}
\bibliography{thf-dissoc}

\end{document}